\documentclass[12pt]{article}
\usepackage{epsf}
\usepackage{amsfonts}
\usepackage{amsmath}
\usepackage[latin2]{inputenc}
\usepackage[T1]{fontenc}
\usepackage{epsfig}

\newcommand{\nc}{\newcommand}
\nc{\postscript}[2]
{\setlength{\epsfxsize}{#2\hsize}\centerline{\epsfbox{#1}}}
\nc{\bg}{B. Grzadkowski}
\nc{\non}{\nonumber}
\nc{\hc}{\hbox {H.c.}} \nc{\re}{\hbox {Re}} 
\nc{\mev}{\hbox {MeV}} \nc{\gev}{\;\hbox {GeV}} \nc{\tev}{\;\hbox {TeV}}
\def\lsim{\mathrel{\raise.3ex\hbox{$<$\kern-.75em\lower1ex\hbox{$\sim$}}}}
\def\gsim{\mathrel{\raise.3ex\hbox{$>$\kern-.75em\lower1ex\hbox{$\sim$}}}}

\nc{\prd}[3]{{\it Phys.\ Rev.}\ {{\bf D{#1}} (#2), #3}}
\nc{\prl}[3]{{\it Phys.\ Rev.\ Lett.}\ {{\bf {#1}} (#2), #3}}
\nc{\plb}[3]{{\it Phys.\ Lett.}\ {{\bf B{#1}} (#2), #3}}
\nc{\npb}[3]{{\it Nucl.\ Phys.}\ {{\bf B{#1}} (#2), #3}}
\nc{\ptp}[3]{{\it Prog.\ Theor.\ Phys.}\ {{\bf {#1}} (#2), #3}}
\nc{\zfp}[3]{{\it Z.\ Phys.}\ {{\bf C{#1}} (#2), #3}}
\nc{\epj}[3]{{\it Eur.\ Phys.\ J.}\ {{\bf C{#1}} (#2), #3}}
\nc{\mpla}[3]{{\it Mod.\ Phys.\ Lett.}\ {{\bf A{#1}} (#2), #3}}
\nc{\rmp}[3]{{\it Rev.\ Mod.\ Phys.}\ {{\bf {#1}} (#2), #3}}
\nc{\ijmpa}[3]{{\it Int.\ J.\ of\ Mod.\ Phys.}\
               {{\bf A{#1}} (#2), #3}}
\nc{\Lsp}{\;\;\;\;\;\;\;\;\;\;}  \nc{\LLLsp}{\lspace \lspace}
\nc{\lsp}{\;\;\;\;\;\;}
\nc{\spac}{\;\;\;}
\nc{\noi}{\noindent}
\nc{\beq}{\begin{equation}}
\nc{\eeq}{\end{equation}}
\nc{\bea}{\begin{eqnarray}}
\nc{\eea}{\end{eqnarray}}
\nc{\baa}{\begin{array}}
\nc{\eaa}{\end{array}}
\nc{\bit}{\begin{itemize}}
\nc{\eit}{\end{itemize}}
\nc{\ben}{\begin{enumerate}}
\nc{\een}{\end{enumerate}}
\nc{\bce}{\begin{center}}
\nc{\ece}{\end{center}}

\def\sq2{\sqrt{2}}

\def\ph{\varphi}

\def\m4{m^4(\ph)}
\def\mn2{m_n^2}

\def\r0{|R_0|}

\newlength{\dinwidth}
\newlength{\dinmargin}
\DeclareMathAlphabet{\scr}{U}{rsfs}{m}{n}
\begin{document}
\pagestyle{empty}

\noindent
\begin{flushright}
$\vcenter{ \hbox{{\footnotesize CERN-PH-TH/2004-036}}
\hbox{{\footnotesize IFT-2004-08}}
}$
\end{flushright}

\vskip 1cm

\bce
{\Large\bf Electroweak symmetry breaking and\\
radion stabilization in universal extra dimensions}

\vskip 1cm

\renewcommand{\thefootnote}{\alph{footnote})}
{\sc Patrizia BUCCI$^{\:1),\:}$}\footnote{E-mail address: \tt
Patrizia.Bucci@fuw.edu.pl},\ \
{\sc Bohdan GRZADKOWSKI$^{\:1,\;2),\:}$}\footnote{E-mail address:
\tt Bohdan.Grzadkowski@fuw.edu.pl},\ \

\vskip 0.15cm
{\sc Zygmunt LALAK$^{\:1),\:}$}\footnote{E-mail address:
\tt Zygmunt.Lalak@fuw.edu.pl}\ and\
{\sc Rados\l aw MATYSZKIEWICZ$^{\:1),\:}$}\footnote{E-mail
address: \tt Radoslaw.Matyszkiewicz@fuw.edu.pl}

\vspace*{0.8cm}
{\sl $1)$ Institute of Theoretical Physics,\ Warsaw University\\
Ho\.za 69, PL-00-681 Warsaw, Poland\\
\vspace*{0.4cm}
\sl $2)$
CERN, Department of Physics,\\
Theory Division\\
1211 Geneva 23, Switzerland}

\vspace*{0.7cm}
\centerline{ABSTRACT}
\vspace*{0.5cm}
\ece

We discuss the stabilization of the scalar sector, including the
radion, in the gauge model with one universal extra dimension,
within Higgs and Higgsless scenarios. The stabilization occurs at
the one-loop level, through the fermionic contribution to the
effective potential; in the Higgs case, for stabilization  to take
place the bosonic contribution must be balanced by the fermionic
one, hence the scales of these two cannot differ too much.
However, there is no need for (softly broken) supersymmetry to
achieve the stabilization - it can be arranged for a reasonably
wide range of couplings and mass scales. The primary instability
in the model is the run-away of the radion vacuum expectation
value. It turns out that the requirement of the radion stability,
in the Higgs case, favours a Higgs boson mass below $0.26$~TeV,
which is consistent with the Standard Model upper bound that
follows from the electroweak precision measurements. The typical
radion mass is of the order of $\sim10^{-6}$~eV. The radion mass
can be made larger by rising the scale of fermion masses, as
clearly seen in the Higgsless case. The cosmological constant may
be cancelled by suitable counterterms, in such a way that the
stabilization is not affected.

\vspace{0.6cm}

\vfill



\newpage
\renewcommand{\thefootnote}{$\sharp$\arabic{footnote}}
\pagestyle{plain} \setcounter{footnote}{0}
\section{Introduction}
Understanding the origin of the electroweak symmetry breaking at $M_{EWB} \leq 1$ TeV and the fermion mass
generation appears to be one of the big theoretical challenges of contemporary physics. There exist various
ideas of how to give mass to the gauge bosons mediating weak interactions and how to, simultaneously,
render the scale of the breaking in the $1$ TeV range in the presence of radiative corrections.
One of the most natural tools is supersymmetry,
another one - extra dimensions, which offer new possibilities both for electroweak breaking and
for supersymmetry breaking (e.g. by suitable boundary conditions).
However, with extra dimensions there appears a new issue in the game - the
problem of stabilization of compact dimensions, which, in fact, seems to be a disguised version
of the familiar, well-known hierarchy problem. In this note we would like to
address, in the simplest possible set-up, the question of the interrelation
between these issues. About the supersymmetry breaking we shall be rather
brief here, simply assuming that it is somehow broken, perhaps even in
a hard way; hence, for instance, the number of fermions does not need to match
the number of bosons in the model. Taking that for granted, we consider here one-loop corrections
to the effective potential for the radion (which is the scalar excitation of the
extra-dimensional metric tensor whose
vacuum expectation value fixes the size of extra dimensions)
and the Higgs boson, as a
source of the radion stabilization. It turns out that it is possible to create
a non-trivial and quasi-realistic minimum of the effective potential in the space spanned
by the scalar fields of the theory, with one universal extra dimension
in two cases of special interest. Firstly, when the electroweak breaking is
caused by the condensation of the higher-dimensional Higgs scalar, and
secondly in the Higgsless case, when we imagine that the massless mode of gauge fields is
removed from the spectrum by boundary conditions.
The notorious feature of the set-up containing a Higgs boson is a light, in fact too light,
radion excitation. In the Higgsless case it is much easier to avoid such a problem: it
is possible to raise the radion mass by coupling a radion to heavy
fermions living in the bulk of the model. We find it amusing and encouraging at the same time to find
stable vacuum states with stable extra dimensions and broken gauge symmetry
with essentially arbitrarily broken supersymmetry.

The paper is organized as follows. In Section~\ref{set-up} we define the 5d Higgs model. Section~\ref{comp}
contains details of the reduction from 5d to 4d.
In Section~\ref{radcor} we calculate the one-loop effective potential in order to determine the radion mass
and we comment on the existing experimental constraints on the radion mass.
Section~\ref{hless} is devoted to the discussion of the radion stability
in the Higgsless scenario. Summary and comment on consequences of possible variations
of the set-up adopted here are presented in Section~\ref{sum}.
The appendix contains details of the derivation of the effective potential.

\section{General set-up}
\label{set-up}

Let us start with the following action in 5d
\begin{equation}
     S=S_g+S_s+S_f+S_v+S_{gf}\,,
\end{equation}
where $S_g$ denotes the Einstein--Hilbert action,
\begin{equation} \label{actiongrav}
      S_g=-\frac{1}{2}M_5^{3}\int_{0}^Ldy\int d^{4}x\sqrt{G}R^{(5)}\,,
\end{equation}
where $R^{(5)}$ is the Ricci scalar constructed from the 5d metric tensor $G_{MN}$
and $L=2\pi \rho$.
The scale $M_5$ sets the 5d gravitational coupling. The notation for the Lorentz indices
and space-time coordinates is
as follows: $M,N=0,1,2,3,5$;
$\mu,\nu=0, 1, 2,3$ and $y=x^{5}$ is the coordinate of the extra dimension.
The action for the complex scalar field and vector bosons reads
\bea
S_s&=&\int_{0}^Ldy\int d^{4}x\sqrt{G}\left[(D_M\phi)^*(D^M\phi)-V^{(5)}(\phi)\right]\,,\\
S_v+S_{gf}&=&\int_{0}^Ldy\int d^{4}x\sqrt{G}\left\{ -\frac{1}{4}F^{MN}F_{MN}
-\frac{1}{2\xi}\left[\partial_{\mu}A^{\mu}-\xi\left(\partial_{5}A_{5}+
e v \chi \right)\right]^{2} \right\}\,,
\eea
where $v$ is the vacuum expectation value  of the zero mode of the scalar field,
$e_4\equiv e_5/\sqrt{L}$ will appear to be
the effective 4d electromagnetic gauge coupling and
\begin{eqnarray}
&F_{MN} =\partial_M A_N-\partial_N A_M \,, & D_M =\partial_M+{\rm i}e_5 A_M \,,\nonumber\\
&V^{(5)}(\phi)=\lambda_5\left(|\phi|^{2}-\frac{\mu^{2}}{2\lambda_5}\right)^{2}
\,, & \phi =\frac{1}{\sqrt{2}}\left(h+{\rm
i}\chi\right)\,.\nonumber
\end{eqnarray}
Here we will adopt
the Landau gauge, which is equivalent to the limit  $\xi\to 0$.

In order to construct a Standard Model-like theory, we will follow
Ref.~\cite{Bucci:2003fk} and introduce two fermionic fields
$\psi=\psi(x,y)$ (charged) and $\lambda=\lambda(x,y)$ (neutral)
with the following transformation properties: \beq \psi(x,y) \to
\gamma_5 \psi(x,L-y)\lsp {\rm and} \lsp \lambda(x,y) \to -\gamma_5
\lambda(x,L-y)\,, \label{sym_f} \eeq while for the bosonic fields
we assume \beq \phi(x,y)\to \phi(x,L-y),\;\; A_\mu(x,y)\to
A_\mu(x,L-y) \;\; {\rm and} \;\; A_5(x,y) \to -A_5(x,L-y)\,.
\label{sym_s} \eeq Then, the invariant fermionic action reads:
\beq S_f=\int_{0}^Ldy\int d^{4}x\sqrt{G}\left[{\rm
i}\overline{\psi}\gamma^M(\partial_M+{\rm i}e_5A_M)\psi+ {\rm
i}\overline{\lambda}\gamma^M\partial_M\lambda-
\left(g_5\overline{\psi}\phi\lambda+\hc\right)\right]\,,
\label{feract} \eeq As can be seen, the fermion mass term is
generated (as in the Standard Model (SM)) by the scalar vacuum
expectation value.

The size of the extra dimension $L=2\pi\rho$ is an arbitrary parameter with the dimension of length.
It has no physical meaning. What is physically meaningful is the
distance along the compact dimension
\begin{equation}
L_{\rm phys}=\int_{0}^Ldx^{5} \sqrt{-G_{55}}\,.
\end{equation}

\section{Compactification}
\label{comp}

Let us construct the 4d effective theory.
Since hereafter we will consider neither Kaluza--Klein (KK) modes of the 4d metric $g_{\mu\nu}(x,y)$
nor those of the radion $R_0(x,y)$, the background 5d metric can be parametrized as
\begin{equation}
      G_{MN}=\left(\begin{array}{cc}g_{\mu\nu}(x)&0\\0&-R_{0}^{2}(x)\end{array}\right)\,.
\label{metric}
\end{equation}
The compactification of the extra
dimension is specified by the following $S^{1}/Z_{2}$ orbifold
conditions:
\begin{eqnarray}
& A_\mu(x,y)=A_\mu(x,-y)\,, \;\;\; & A_5(x,y)=-A_5(x,-y)\,,\nonumber\\
&\phi(x,y)=\phi(x,-y)\,,&\nonumber\\
&\psi_R(x,y)=\psi_R(x,-y)\,, \;\;\; & \psi_L(x,y)=-\psi_L(x,-y)\,,\nonumber\\
&\lambda_L(x,y)=\lambda_L(x,-y)\,, \;\;\;
&\lambda_R(x,y)=-\lambda_R(x,-y)\nonumber\,. \label{orbi}
\end{eqnarray}
Moreover, the fields should remain unchanged under the shift $y\rightarrow y+2 \pi \rho$.
The resulting KK expansion is given in the appendix.

An important remark is in order here. In general, instead of discussing the circle and its
symmetries, one could go immediately to a line segment and impose boundary conditions on the fields
by coupling them to suitable sources localized on the branes. These sources appear in the equations of motion
and enforce a definite behaviour of the fields at the boundaries.
This way one may obtain boundary conditions corresponding to fields living on a quarter of a circle, or on
$S^{1}/ Z_2 \times Z^{'}_2$. It is often convenient to discuss such set-ups on a circle; however one then has
to accept fields that are not periodic. We shall discuss such a case later in the paper.

After integrating out the extra coordinate, we find the effective Einstein--Hilbert term multiplied by  a
a power of the radion field
       \begin{equation} \label{ncanein}
      S_g^{\rm eff}=-\frac{1}{2}M_5^{3}2\pi \rho\int d^{4}x\sqrt{-g}|R_0|R^{(4)} \,.
      \end{equation}
It will be useful to transform the above action to the Einstein frame by performing the following Weyl rescaling
      \begin{equation} \label{ncaneinn}
      g_{\mu\nu}\longrightarrow |R_0|^{-1}g_{\mu\nu}\,,
      \end{equation}
      which results in the gravitational action
       \begin{equation}
      S_g^{\rm eff}=-\frac{1}{2}M_4^{2}\int d^{4}x\sqrt{-g}R^{(4)}+\frac{1}{2}\int d^{4}x\sqrt{-g}\partial_{\mu}r
\partial^{\mu}r \,, \label{eaction}
      \end{equation}
       where we have defined  $r=\sqrt{3/2}M_4\log{|R_{0}|}$, and $M_4$ denotes the effective 4d
Planck scale.
It is worth emphasizing that the Weyl rescaling is essential here;
it is necessary to properly identify the 4d metric
as the one that appears on the r.h.s. of Eq.~(\ref{ncaneinn}), for which we obtain the canonical form
of the Einstein gravity action in Eq.~(\ref{eaction}).
Notice also that our definition $M_4^{2}=2\pi\rho M_5^{3}$ does not express a relation between four-
and five-dimensional Planck scales.
Precise analysis of the Newton law shows that the actual 5d
Planck scale,
related to the 5d Newton constant, reads  $M_5^{\rm true}=M_5\langle |R_{0}|\rangle^{-1/3}$, and then
$M_4^{2}=2\pi\rho \langle |R_{0}|\rangle (M_5^{\rm true})^{3}$.

It is straightforward to verify that, after the Weyl rescaling, we must also
rescale $A_\mu$ and $A_5$ to obtain canonical
kinetic action
      \begin{eqnarray}
      A_{\mu}\longrightarrow |R_{0}|^{-\frac{1}{2}} A_{\mu}\,,\quad A_{5}\longrightarrow |R_{0}| A_{5}\,.
\label{vecres}
\end{eqnarray}
The following redefinition is necessary for fermionic fields as well
    \begin{eqnarray}
      \psi\longrightarrow |R_{0}|^{\frac{1}{4}} \psi\,,\quad \lambda\longrightarrow
      |R_{0}|^{\frac{1}{4}}\lambda\,.
\label{ferres}
\end{eqnarray}
After the Weyl rescaling the 5d metric takes  the following form
\begin{equation}
ds^2=R_{0}^{-1}g_{\mu \nu}dx^{\mu}dx^{\nu}-R_{0}^2dy^2\,. \label{5dmet}
\end{equation}
Hence, the physical size of the extra dimension is given  by $L_{\rm phys}=2 \pi \rho \langle R_{0}\rangle$,
where $\langle R_{0}\rangle$ is  determined by the quantum corrections computed by expanding the Lagrangian
around the classical solution of the 5d Einstein equations\footnote{Note that in Eq.~(\ref{5dmet}),
it is $g_{\mu\nu}$ which is the 4d metric.}
\begin{equation}
\begin{array}{c}
g_{\mu \nu}=\eta_{\mu \nu}\\
\,\,\, \\
R_{0}(x)=\langle R_{0}\rangle={\rm const.}\,,
\end{array}
\end{equation}
$\eta_{\mu \nu}$ being the Minkowski metric.

It should be noted that the rescaling (\ref{vecres}) and (\ref{ferres}) generates a number of
derivative-type couplings of the radion. Those, however, are not relevant to the calculation of the one-loop
effective potential and therefore will not longer be considered.

\section{Radiative corrections}
\label{radcor}

After the Weyl rescaling, the 4d tree-level potential in the Landau gauge is obtained,
as the following integral over the extra dimension:
\begin{equation}
V^{4}(\phi,r)=\int_{0}^{2 \pi \rho}dy e^{-\alpha  r} \left( V^{5}(\phi)+e^{-2 \alpha  r}|D_5\phi|^2\right)\,,
\end{equation}
where
\beq
\alpha=\frac{\sqrt{2}}{\sqrt{3}M_4}\,.
\eeq
Here we will consider only the case in which the zero-mode\footnote{For an example of a  model with
a non-trivial profile of the Higgs background field, which means non-zero vacuum expectation values
for KK modes,
see \cite{Grzadkowski:2004mg}.}
of the real component $h(x)$ of the scalar field $\phi(x,y)$ and possibly the radion $r(x)$ can
acquire vacuum expectation values.

Following Ref.~\cite{Bucci:2003fk}, we shall adopt, to
compute the contribution of the KK tower to the effective potential, the
regularization scheme worked out by Delgado et al. (DPQ, see \cite{Delgado:1998qr}, see also \cite{Antoniadis:1990ew}, \cite{Antoniadis:1998sd} for earlier results).
The result of Ref.~\cite{Bucci:2003fk} was obtained in the absence of gravity, for a flat metric, and
assuming that the radion was stabilized. What we are studying now is the possibility to stabilize the
radion through electroweak radiative corrections and at the same time to reproduce the usual 4d SM-like theory.
We will start from formula (\ref{dpq}), which for the scalar field is given by:
\begin{equation}
V_{\rm 1-loop}=\frac{1}{2}\sum_{0}^{\infty}\int\frac{d^{4}p}{(2 \pi)^4}\log{ \left[ (p^2+m^2_{h_{0}})+
{n^2 \over \rho^2} \, e^{-3 \alpha \langle r \rangle}\right] }\,,
\end{equation}
which is equal to:
\begin{equation}
V_{\rm 1-loop}=\frac{1}{2}\sum_{0}^{\infty}\int\frac{d^{4}p}{(2 \pi)^4}
\log{ \left[ l^2(p^2+m^2_{h_{0}})e^{3 \alpha \langle r \rangle}+n^2 \pi^2 \right] }-
\frac{3 \alpha}{2}\sum_{0}^{\infty}\int\frac{d^{4}p}{(2 \pi)^4} \langle r \rangle\,,
\end{equation}
where $l=\pi \rho$.
The second term above is a divergent contribution that vanishes
in the dimensional regularization. Applying the DPQ procedure (see the appendix for details)
we obtain the following contribution from
the KK tower of non-zero modes of the scalar $h$:
\begin{eqnarray}
V^{(\infty)}_{h}&=& e^{\frac{3}{2}\alpha \langle r\rangle}\frac{\rho }{60\pi}|m_{h_{0}}|^{5}\,,\nonumber\\
V^{(R)}_{h}&=&-e^{-6\alpha \langle
r\rangle}\frac{1}{64\pi^{6}\rho^{4}}\left[x_{h}^{2}{\rm
Li_{3}}\left(e^{-x_{h}}\right)+3 x_{h}{\rm
Li_{4}}\left(e^{-x_{h}}\right)+3{\rm
Li_{5}}\left(e^{-x_{h}}\right)\right]\,,
\end{eqnarray}
where $x_{h}=e^{\frac{3}{2}\alpha \langle
r\rangle}2\pi\rho|m_{h_{0}}|$,  and scalar masses are defined in Eq.~(\ref{mass}).
The result is in agreement with that of Ref.~\cite{Ponton:2001hq}.

In the case of a mixing, as in  the system ($\chi, A_{5})$, we use the  procedure described in
\cite{Bucci:2003fk} to obtain one-loop corrections
\begin{eqnarray}
V^{(\infty)}_{\rm mix}&=& -e^{\frac{3}{2}\alpha \langle
r\rangle}\frac{\rho
}{32}b^{\frac{1}{4}}\left(b-\frac{a^{2}}{4}\right)
{\rm F}\left(-\frac{1}{4},\frac{7}{4};2;1-\frac{a^{2}}{4b}\right)\,,\nonumber\\
V^{(R)}_{\rm mix}&=&-e^{-\frac{3}{4}\alpha \langle
r\rangle}\frac{b^{\frac{3}{4}}
\left(2\sqrt{b}+a\right)^{\frac{1}{4}}}{16\pi^{2}\sqrt{\rho}}{\rm
Li_{\frac{3}{2}}} \left[\exp\left( \left( -2\pi\rho
e^{\frac{3}{2}\alpha \langle r\rangle}\right) \left(
2\sqrt{b}+a\right)^{\frac{1}{2}}\right) \right]\,,
\end{eqnarray}
where we have defined
\begin{eqnarray}
a&=& e^{-\alpha \langle r\rangle}\left(e_{4}^{2}\langle h\rangle^{2}-\mu^{2}+
\lambda_{4}\langle h\rangle^{2}\right) \,,\nonumber\\
b&=& e^{-2\alpha \langle r\rangle}e_{4}^{2}\langle
h\rangle^{2}\left(-\mu^{2}+\lambda_{4}\langle
h\rangle^{2}\right)\,.
\end{eqnarray}
{}From Eq.~(\ref{mass}) one can see that the radion
mixes only with the zero mode of the scalar field $h_{0}$. We can
calculate the  eigenvalues of the squared mass matrix for these
fields
\begin{equation}
m^{2}_{1,2}=\frac{1}{2}\left(m_{h_0}^{2}+m_{r}^{2}\pm\sqrt{(m_{h_0}-m_{r})^{2}+4m_{r\;
h_{0}}^{2}}\right)\,.
\end{equation}
The  contribution to the one-loop potential from a single
scalar field is
\begin{equation}
V^{0}_{s}=\frac{1}{64\pi^{2}}m_{s}^{4}\left[\log\left(\frac{m_{s}^{2}}{\kappa^{2}}\right)-\frac{3}{2}\right]\,,
\end{equation}
where $\kappa$ denotes the renormalization scale.
Therefore, the total contribution of the scalar fields to the one-loop effective potential is given by
\begin{equation}
V^{\rm
1-loop}_{s}=\frac{1}{2}\left(V^{(\infty)}_h+V^{(R)}_{h}-V^{0}_{h}+2V^{0}_{1}+2V^{0}_{2}+
V^{(\infty)}_{\rm mix}+V^{(R)}_{\rm
mix}+V^{0}_{\chi}-V^{0}_{A_5}\right)\,.
\end{equation}

Let us find the contributions to the effective potential coming from the other fields.

For the vector boson, the DPQ procedure leads to
\begin{eqnarray}
V^{(\infty)}_{A_{\mu}}&=& e^{\frac{3}{2}\alpha \langle r\rangle}\frac{\rho }{60\pi}|m_{A_{\mu 0}}|^{5}\,,\nonumber\\
V^{(R)}_{A_{\mu}}&=&-e^{-6\alpha \langle
r\rangle}\frac{1}{64\pi^{6}\rho^{4}}\left[ x_{A_{\mu}}^{2}{\rm
Li_{3}}\left(e^{-x_{A_{\mu}}}\right)+3 x_{A_{\mu}}{\rm
Li_{4}}\left(e^{-x_{A_{\mu}}}\right)+ 3{\rm
Li_{5}}\left(e^{-x_{A_{\mu}}}\right)\right]\,,
\end{eqnarray}
where $x_{A_{\mu}}=e^{\frac{3}{2}\alpha \langle
r\rangle}2\pi\rho|m_{A_{\mu 0}}|$ and the vector masses are defined in
Eq.~(\ref{massvb}). The total contribution of the vector fields to
the one-loop effective potential is
\begin{equation}
V^{\rm
1-loop}_{v}=\frac{3}{2}\left(V^{(\infty)}_{A_{\mu}}+V^{(R)}_{A_{\mu}}+V^{0}_{A_{\mu}}\right)\,,
\end{equation}
where
\begin{equation}
V^{0}_{A_{\mu 0}}=\frac{1}{64\pi^{2}}m_{A_{\mu
0}}^{4}\left[\log\left(\frac{m_{A_{\mu
0}}^{2}}{\kappa^{2}}\right)-\frac{5}{6}\right]\,.
\end{equation}

The fermionic contributions to the one-loop effective potential are
\begin{equation}
V^{\rm 1-loop}_{f}=-4\left(V^{(\infty)}_{f}+V^{(R)}_{f}\right)\,,
\end{equation}
where
\begin{eqnarray} \label{hmf}
V^{(\infty)}_{f}&=& e^{\frac{3}{2}\alpha \langle r\rangle}\frac{\rho }{60\pi}|m_{f}|^{5}\,,\nonumber\\
V^{(R)}_{f}&=&-e^{-6\alpha \langle
r\rangle}\frac{1}{64\pi^{6}\rho^{4}}\left[x_{f}^{2}{\rm Li_{3}}
\left(e^{-x_{f}}\right)+3 x_{f}{\rm
Li_{4}}\left(e^{-x_{f}}\right)+3{\rm
Li_{5}}\left(e^{-x_{f}}\right)\right]\,.
\end{eqnarray}
We have defined $m_{f}=e^{-\frac{1}{2}\alpha \langle
r\rangle}\frac{g_{4}}{\sqrt{2}}\langle h\rangle$ and
$x_{f}=e^{\frac{3}{2}\alpha \langle r\rangle}2\pi\rho|m_{f}|$. The
mass $m_f$ comes from the diagonalization of the fermion masses,
which are written in Eq.~(\ref{massf}). The total one-loop
potential, including all contributions, takes the form
\begin{equation}
V^{\rm 1-loop}_{\rm tot}=V^{\rm 1-loop}_{s}+V^{\rm
1-loop}_{v}+V^{\rm 1-loop}_{f}+V^{\rm tree}_{s}\,,
\end{equation}
where
\begin{equation}
V^{\rm tree}_{s}=e^{-\alpha \langle
r\rangle}\left(-\frac{1}{2}\mu^{2}\langle h\rangle^{2}+
\frac{1}{4}\lambda_{4}\langle
h\rangle^{4}+\frac{\mu^{4}}{4\lambda_4}\right)\,.
\end{equation}
This effective potential has been obtained by neglecting some diagrams. More precisely, the missing ones are those involving
virtual fluctuations of the 5d metric and their KK excitations.
It is easy to see that these diagrams can be safely neglected here.
The general argument consists of the observation that, in order to turn the fluctuations
of the 5d metric, $h_{MN} $,
into canonical dimensionful fields in 4d, one needs to multiply them by the
4d Planck scale, $h_{MN} \rightarrow h_{MN} M_4$, which means that their couplings to matter are
suppressed by inverse powers of $M_4$.
Hence, generally, it is justified to neglect these metric fields in the loops
as long as one finds stabilization due to the matter loops. This point is well
illustrated by considering radion loops.
In this case the most important diagrams are those involving a single radion internal line, which is quadratically
divergent, and a loop made of a radion line and a scalar line, which is logarithmically divergent. These diagrams give
a contribution to the effective potential of the order of
\begin{equation}
\alpha^2\Lambda^2\sim \left(\frac{\Lambda}{M_{4}}\right)^2\,.
\end{equation}
Since we expect that the physical cut-off for the 4d physics is
much smaller\footnote{Note that, if the Higgs-boson mass is small enough, the electroweak vacuum of the
one-loop effective potential is unstable, the cutoff that follows could be as small as a few TeV.}
than the Planck scale, we could therefore have left out these
contributions while retaining the ones previously discussed, due to scalars, fermion and the gauge fields.
The approximation adopted here consistently treats the gravitational interactions at the classical level, while
the crucial quantum effects (including the non-zero vacuum expectation value for the radion)
emerge from the matter fields.

\begin{figure}[h]
        \begin{center}
        \epsfig{file=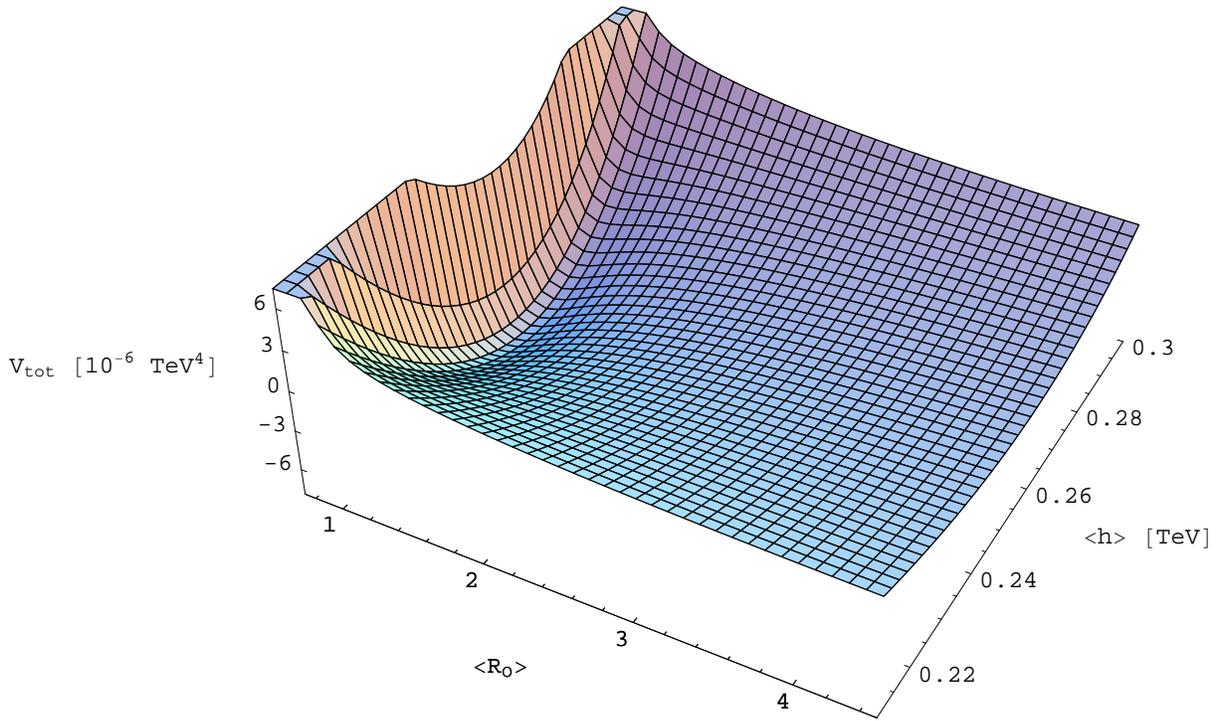, width=.95 \linewidth}
        \caption{\small{The total potential for the parameters given in (\ref{valueforpar}). The minimum appears
        at the point $( \langle R_0 \rangle,\langle h\rangle)=(1,\;0.258{\rm~TeV})$. } \label{w1}}
        \end{center}
\end{figure}

The total effective potential has been analysed as a function of two parameters:
$\langle h\rangle$ and $\langle r\rangle$.
Numerical calculations have been performed for the following set of parameters:
\begin{eqnarray}\label{valueforpar}
&\mu=\frac{m_H}{\sqrt{2}}\,,\quad
\lambda_4=\frac{1}{2}\left(\frac{m_H}{0.246 {\rm
~TeV}}\right)^{2}\,,
\quad m_H=0.12 {\rm ~TeV}\,, \nonumber\\
& \kappa=0.1{\rm ~TeV}\,,\quad M_4=2\times 10^{15}{\rm ~TeV}, \quad m_{t}=0.175{\rm ~TeV} \,,\nonumber\\
&g_4=\frac{\sqrt{2}m_{t}}{0.246 {\rm ~TeV}}\,,\quad
e_4=\sqrt{4\pi/137}\,,\quad\rho=2.11 {\rm ~TeV}^{-1}\,,
\end{eqnarray}
where $m_H$ denotes the tree-level mass of the Higgs boson.
We have obtained the  minimum of the $V^{\rm 1-loop}_{\rm tot}(\langle r\rangle,\langle h\rangle)$
at the point $(\langle r\rangle,\langle h\rangle)
=(1.89 \times 10^{8},0.259)$~TeV. Notice that this result corresponds to  $\langle R_{0}\rangle =1$ (see Fig.~\ref{w1}).
We can also compute effective masses$^2$ for the radion-$h_0$ system:
\begin{eqnarray}
&&m^{2}_{h_{0}}=\frac{\partial^{2}V^{\rm 1-loop}_{\rm
tot}}{\partial \langle h\rangle^{2} }\bigg|_{\rm min} =0.014 {\rm
~TeV}^{2} \,,\nonumber\\ && m^{2}_{r}=\frac{\partial^{2}
V^{1-loop}_{\rm tot}}{\partial \langle r\rangle^{2} }\bigg|_{\rm
min}=1.14\times 10^{-35} {\rm ~TeV}^{2} \,,
\nonumber\\&&m^{2}_{r\; h_{0}}=\frac{\partial^{2} V^{\rm
1-loop}_{\rm tot}}{\partial \langle r\rangle\partial \langle
h\rangle }\bigg|_{\rm min}=3.38 \times 10^{-20} {\rm ~TeV}^{2}\,.
\end{eqnarray}
We have always chosen the parameter $\rho$ in such a way that the minimum of the complete  potential appears
at the point $\langle r\rangle\ll M_{\rm Pl}$, which implies $R_0\approx 1$ (see discussion below).
In such a case the physical radius of the fifth dimension is given by the
parameter  $\rho$.
\begin{table}[h]
\begin{center}
\begin{tabular}{|c|c|c|c|c|c|}\hline
$m_H$ [TeV]&$m_h$ [TeV]&$m_{r}$ [$10^{-6}$~eV]&$\rho^{-1}$ [TeV] & $\langle h\rangle$ [TeV]&
$\Lambda_4$ [$10^{-6}$~TeV]\\\hline
              0.10&0.098&3.5&0.49&0.265&$-7.90$\\\hline
              0.12&0.119&3.4&0.47&0.259&$-7.59$\\\hline
              0.14&0.139&3.3&0.47&0.255&$-7.36$\\\hline
              0.16&0.160&3.2&0.46&0.252&$-7.09$\\\hline
              0.18&0.181&3.0&0.44&0.250&$-6.68$\\\hline
              0.20&0.202&2.8&0.42&0.249&$-6.05$\\\hline
              0.22&0.225&2.4&0.39&0.247&$-5.06$\\\hline
\end{tabular}
\caption{Higgs boson and radion mass, the scale of the extra dimension $\rho^{-1}$, the Higgs-boson
vacuum expectation values and the resulting cosmological constant,  obtained for different input tree-level Higgs boson
masses $m_H$.}
\label{tab1}
\end{center}
\end{table}
Let us explain the way we adjust the $\rho$, which parametrizes the physical masses and couplings in 4d.
The point is that we are interested in a specific range of the effective physical scales as seen in 4d,
which we consider realistic. However, these physical scales depend on the expectation value of the radion,
which we need to determine dynamically, and this dependence occurs through the factors that are powers of
$e^{\alpha \langle r \rangle}$ (the parameters that are radion-independent are those that define the
Lagrangian in 5d). To be able to follow the dynamical determination of the radion, we define
auxiliary 4d parameters, which are radion-independent: $\lambda_4, \, e_4, \, g_4$ differ from the
physical ones by the suitable powers of  $e^{\alpha \langle r \rangle}$ and we note that
the auxiliary parameters are equal to physical ones at $\langle R_0 \rangle = e^{\alpha \langle r \rangle} \approx 1$.
The usual approach would be to fix $\rho$,
which sets the physical scale of the fifth dimension, to some convenient value, e.g. $\rho=1$,
and to keep it constant during the
calculations. However, the above
reasoning suggests that the opposite is more convenient: in each model under discussion we shall fix the
expectation value of the radion to be
equal to unity, and achieve this by changing the value of $\rho$. It is obvious that physically this is
a legal point of view. In 5d
the meaningful quantity is in fact $2\pi \rho R_0$, and any change of $\langle R_0 \rangle$ can be compensated by
an adjustment of $\rho$, while keeping
the parameters of the Lagrangian, hence the 5d model, unchanged.
When we compactify and switch over to the 4d language the situation becomes slightly more complicated, since
the 4d couplings, say $\lambda_4$,
are related to the 5d ones by a power of $\rho$. Hence to stay in the same 5d model one would have to change
$\lambda_4$ together with $\rho$.
This is not what we want to do: we are interested in 4d models and keep 4d couplings constant. This is
perfectly acceptable from the point of view of the
4d physics; one only needs to keep in mind that in the present picture different values of $\rho$ correspond
to slightly different 5d couplings, hence
slightly different 5d models.
Having said this, let us define the procedure that brings us down to $\langle R_0 \rangle=1$ and allows us
to identify the physical masses in 4d in a straightforward manner.
We start with an arbitrarily chosen value of $\rho$ and minimize the one-loop
potential to find $\langle R_0 \rangle$.
Then we repeat the procedure, taking $\rho^{(1)}= \langle R_0 \rangle \rho$. Then we repeat the steps again
and again until
we reach $\langle R_0 \rangle \approx 1$, taking for each consecutive iteration
$\rho^{(n)} = \langle R_0 \rangle^{(n-1)} \rho^{(n-1)}$, where $n$ denotes
the parameter of the $n$-th iteration. The procedure converges to
$\langle R_0 \rangle \approx 1$ within just a few iterations (as expected, since the physical size
of the fifth dimension is $ 2 \pi \rho \langle R_0 \rangle$).

In addition,  the calculations have been done for  various values
of the Higgs mass, and the results are listed in Table~\ref{tab1}.
We have chosen the tree-level Higgs boson mass to be in agreement
with the electroweak measurements, i.e. roughly between $0.1$~TeV
and $0.22$~TeV. It turns out that for $m_H \gsim 0.26$~TeV the
effective potential becomes unstable: the radion vacuum
expectation value runs away to infinity. It is amusing to notice
that $m_H =  0.204$~TeV is the electroweak 95\% CL upper bound on
the SM Higgs boson mass~\cite{Hagiwara:fs}. The existence of the
minimum is a result of an interplay between bosonic and fermionic
contributions to the effective potential, so the largest Higgs
mass for which we obtain stability is correlated with the top
quark mass $m_t=0.175$~TeV. Therefore even though our toy model
does not reflect all the features of the real 5d SM, it does
nevertheless contain right mass scales. At the same time, it seems
to  favour the range of the Higgs boson mass that is also
anticipated by the one-loop predictions of the SM. We find this
nice agreement quite amusing. Note also in the table that the
diagonal Higgs boson mass $m_h$ and the vacuum expectation values
$\langle h \rangle$ are almost insensitive to the input,
tree-level Higgs boson mass $m_H$; this is an obvious result of
the very small mixing with the radion. The diagonal radion mass
varies between $3.47\times 10^{-6}$ and $2.39\times 10^{-6}$~eV.
The resulting size of the extra dimension, $\rho^{-1}\simeq
0.4$~--~$0.5$~TeV, roughly agrees with the existing bound on the
size of one universal extra dimension~\cite{Appelquist:2000nn}.

An important comment is in order here. From the 5d
point of view the meaningful physical quantity is the physical size of the fifth dimension, which
is given by $ L_{\rm phys}=2 \pi \rho \langle R_0 \rangle$. Hence, at first sight,
in various physical quantities the powers of $\rho$ should always multiply the same powers of $R_0$.
However, this is not the case in the 4d Lagrangian and consequently,
one finds in the effective potential an extra $R_0$ dependence that is not of the form $2\pi\rho R_0$.
A closer inspection of the effective potential shows that
this extra dependence on $\langle R_0 \rangle$ has its roots
in one additional power of $R_0$, which shows up in mass terms (both in those that originate
from the $\mu^2$ scalar mass term and also in KK mass terms) in the 4d Lagrangian.
However, this is correct and the reason  can be seen in Eqs.~(\ref{ncanein}) and (\ref{ncaneinn}).
The point is that in (\ref{ncanein}) we have chosen to perform the Weyl rescaling using only the $R_0$,
while $2\pi\rho$ becomes swallowed by the definition of the 4d Planck scale
$M^{2}_4 = 2 \pi \rho M^{3}_5$.

We have seen above that the radion field turns out to be very light and that it experiences a negligible mixing
with the Higgs field:
for $M_{h}=0.12$~TeV, we obtain $m_{r}=3.4\times 10^{-6}$~eV. Such a light field can modify the Newtonian gravity.
A particle of mass $\sim 3.4\times 10^{-6} $ eV can mediate forces over a range of $\sim 37 $~cm,
see Ref.~(\cite{Chacko:2002sb}).
Therefore such a small radion  mass is excluded by experiments.
It is possible to increase the radion mass by one or two orders of magnitude by
rising the fermion masses (see also the next section), but heavy radion is not a
natural phenomenon within the present set-up.
As pointed out in \cite{Chacko:2002sb} the explanation for a such low mass is due
to the higher dimensional general covariance, which forbids a radion mass term in the higher dimensional theory.
Therefore, in the flat 4d theory
the radion mass term can  appear only as a loop effect, and since
the couplings of the radion are Planck-scale-suppressed, the resulting mass is naturally small.

\section{The Higgsless theory}
\label{hless}

In the SM, the Higgs mechanism generates masses
for the fermions and for the vector bosons. In the Higgsless
theory one may assume that the fermion masses emerge from some
additional dynamical mechanism \cite{Csaki:2003dt}--
\cite{Barbieri:2003dd}, e.g. by the
fermion condensation, while the masses of the vector bosons are due to
a global breakdown of the gauge symmetry by boundary conditions imposed
along extra dimensions. This is a noteworthy alternative to the usual Higgs
mechanism, but also a particularly clear limit of the general case
considered in the earlier sections, thus of particular interest to us.

Let us begin the discussion with  the model that does not contain a 5d scalar field
\begin{equation}
     S=S_g+S_f+S_v+S_{gf}\,,
\label{higgsless}
\end{equation}
where $S_g$ denotes the Einstein--Hilbert action (\ref{actiongrav}) and the
action for fermions and vector boson is given by
       \begin{equation}
      S_f=\int_{0}^Ldy\int d^{4}x\sqrt{G}\left[{\rm i}\overline{\psi}\gamma^M
      (\partial_M-{\rm i}e_5A_M)\psi+{\rm i}\overline{\lambda}\gamma^M\partial_M\lambda
      - \left(m_5\overline{\psi}\lambda+\hc\right)\right]
      \end{equation}
       and
       \begin{equation}
S_v+S_{gf}=\int_{0}^Ldy\int d^{4}x\sqrt{G}\left\{ -\frac{1}{4}F^{MN}F_{MN}
-\frac{1}{2\xi}\left[\partial_{\mu}A^{\mu}-\xi\partial_{5}A_{5}\right]^{2} \right\}\,,
      \end{equation}
respectively. The Yukawa interaction present in the Lagrangian (\ref{feract}) has been replaced here
by the 5d mass term $m_5 \overline{\psi}\lambda$.
Here again we require the invariance of the action with respect to the
transformations (\ref{sym_f}) and (\ref{sym_s}), which eliminates the possibility
of diagonal fermionic mass terms.
Here, however, we must modify the set-up employed in Section~\ref{set-up} and assign
the U(1) charge to the fermion $\lambda$ in such a way that the mixed mass term
$\overline{\psi}\lambda$ is gauge-invariant, i.e.
$Q_{U(1)}(\lambda)=Q_{U(1)} (\psi) $ and $\gamma^M\partial_M\lambda
\rightarrow \gamma^M (\partial_M-{\rm i}e_5A_M)\lambda$. Note, however, that
in the one-loop calculation of the effective potential for $r$,
which we will perform here, these two cases, that is invariant and non-invariant
5d fermion mass terms, are indistinguishable and lead to identical conclusions
about the stability of the scalar sector.

To obtain masses for the vector bosons we construct an orbifold $S^1/(Z_2\times Z_2^{'})$
such that the action of the parities on the circle $S^1$ is the following:
$Z_2$: $y\to -y$ and $Z_2^{'}$: $L/2+y\to L/2-y$.
Their action on the field space reads:
\bea
&Z_2: \lsp &A_\mu(x,y)=A_\mu(x,-y), \;\;\; A_5(x,y)=-A_5(x,-y)\label{first}\\
&Z_2^{'}:\lsp &A_\mu(x,L/2+y)=-A_\mu(x,L/2-y), \;\;\;  A_5(x,L/2+y)=A_5(x,L/2-y)\,.
\label{second}
\eea
So, we have assumed $(+,-)$ and $(-,+)$ parities for $A_\mu$ and $A_5$, respectively.
The fermionic boundary conditions remain the same as in the Higgs-like model, i.e.
the right- and left-handed modes transform as $(+,+)$ and $(-,-)$, respectively.
Therefore the fermions are periodic with a period $L$.
The addition of the second requirement (\ref{second})
is a crucial modification of the set-up defined in Section~\ref{set-up}.
This condition causes the breakdown of the gauge symmetry, since
(\ref{first}) alone leads to 4d theory, which is U(1) invariant.
A consequence of (\ref{second}) is that the gauge fields can no longer be periodic; in fact, one finds
that the conditions (\ref{first}) and (\ref{second})
can be consistent only if the gauge fields are antiperiodic:~\footnote{Note that the antiperiodicity
(\ref{aperiod}) is a weaker constraint than (\ref{first}) and (\ref{second}) together.}
    \begin{equation}
A_\mu(x,y+L)=-A_\mu(x,y) \;\;\; A_5(x,y+L)=-A_5(x,y)\,.
\label{aperiod}    \end{equation} This is acceptable as long as
the Lagrangian remains invariant under the twist operator: $T:\;\;
A_M\to -A_M$. Even more, for consistency, the Lagrangian must be
invariant under both $Z_2$ parities acting with respect to each
brane. The symmetry under $Z_2$ is evident. For $Z_2^{'}$,
however, one finds that the interaction between the vector boson
and fermions through the covariant derivative does not fulfil this
requirement, as it is antisymmetric
\begin{equation}
\left[\bar{\psi}\gamma^M e_5
A_M\psi\right](L/2+y)=-\left[\bar{\psi}\gamma^M e_5
A_M\psi\right](L/2-y)\,.
\end{equation}
In order to make the set-up consistent, let us assume that the gauge coupling is odd under
$Z_2^{'}$, so we replace $e_5$ by $\epsilon (y)e_5$ with
\begin{eqnarray}
\epsilon (y) = \left\{
\baa{ccc}
&\vdots &\\
-1&{\rm for}& -3L/2 < y < -L/2\\
+1&{\rm for}& -L/2 < y < L/2\\
-1&{\rm for}& +L/2 < y < 3L/2\\
&\vdots &
\eaa
\right.
\end{eqnarray}
Then the Lagrangian is invariant under both $Z_2$ parities.
Let us note that in the above construction we have not introduced localized
brane terms into the action. As a consequence, each
field which is odd with respect to the given brane must vanish on that brane.
The gauge transformations are not allowed to generate such singular terms,
hence one must require that the gauge variations of the vector bosons do vanish
at the `odd' fixed point. To be more specific
let us consider a gauge transformation with a parameter $\Lambda(x,y)$:
\begin{equation}
\psi\longrightarrow e^{-{\rm i}\epsilon e_5 \Lambda}\psi\,,\quad
\lambda\longrightarrow e^{-{\rm i}\epsilon e_5
\Lambda}\lambda\,,\quad A_M\longrightarrow A_M + \partial_M
\Lambda\,.
\end{equation}
The requirement that such a  gauge transformation does not change parities of
the fields implies that $\Lambda$ is $Z_2$-even with respect to $y=0$ and
$Z_2^{'}$-odd with respect to $y=L/2$.
It is interesting to see that the
above conditions remove the global $U(1)$ transformations from the theory.
This is consistent with the fact that boundary conditions break globally
the group of gauge transformations: not even the global subgroup is left
in the effective 4d model.
Models with jumping gauge couplings were considered before in the
literature, see \cite{Falkowski:2000er}--\cite{Lalak:2003cs}.
Note that after introducing the jumping coupling the observer who travels
around the circle will see precisely the same coupling between the
fermions and the gauge field after passing the brane at $y=L/2$ as before.
Hence the physics on both half-circles remains the same.

Decomposition of the 5d metric tensor  and the KK expansion
      of the fermionic fields is the same as in the previous
      sections. The KK expansion of the vector boson fields reads
       \begin{eqnarray}
A^{\mu}(x,y)&=&\frac{1}{\sqrt{\pi\rho}}
\sum_{n=0}^{\infty}A_{n}^{\mu}(x)\cos\left[y\left(m_n+\frac{\pi}{L}\right)\right] \,,\nonumber\\
A^{5}(x,y)&=&\frac{1}{\sqrt{\pi\rho}}\sum_{n=0}^{\infty}A_{n}^{5}(x)
\sin\left[y\left(m_n+\frac{\pi}{L}\right)\right]\,.
       \end{eqnarray} where $m_{n}=2\pi n/L$.
The following mass terms of the vector bosons are obtained:
\beq
m^{2}_{A_{\mu n}} = e^{-3\alpha \langle
r\rangle}\left(\frac{\pi}{L}+m_{n}\right)^2, \lsp m^2_{A_{5 n}}= 0\,.
\eeq
The scalar modes $A_{5 n}$ are the Goldstons bosons that become longitudinal components of massive
$A_{\mu n}$.

The DPQ procedure leads to
           \begin{eqnarray}
V^{(\infty)}_{A_{\mu}}&=&0\,,\nonumber\\
V^{(R)}_{A_{\mu}}&=&-e^{-6\alpha \langle
r\rangle}\frac{3}{64\pi^{6}\rho^{4}}{\rm Li_{5}}(-1)\,.
           \end{eqnarray}
The total contribution of the vector fields to the one-loop effective potential reads
           \begin{equation}
           V^{\rm 1-loop}_{v}=\frac{3}{2}\left(V^{(\infty)}_{A_{\mu}}+V^{(R)}_{A_{\mu}}\right)\,.
           \end{equation}

The mass terms of the fermions are
         \begin{eqnarray}
m_{\psi_{n}}    &=& -e^{-\frac{3}{2}\alpha \langle r\rangle}m_{n} \,,\nonumber\\
m_{\lambda_{n}} &=& e^{-\frac{3}{2}\alpha \langle r\rangle}m_{n} \,,\nonumber\\
m_{\psi_{n}\;\lambda_{n}}  &=& -e^{-\frac{1}{2}\alpha \langle r\rangle}m_{4}\,,\nonumber\\
m_{\psi_{0R}\;\lambda_{0L}}&=& -e^{-\frac{1}{2}\alpha \langle
r\rangle}m_{4} \,,
       \end{eqnarray}
where $m_{4}=m_{5}/\sqrt{2\pi\rho}$.

\begin{figure}[h]
        \begin{center}
        \epsfig{file=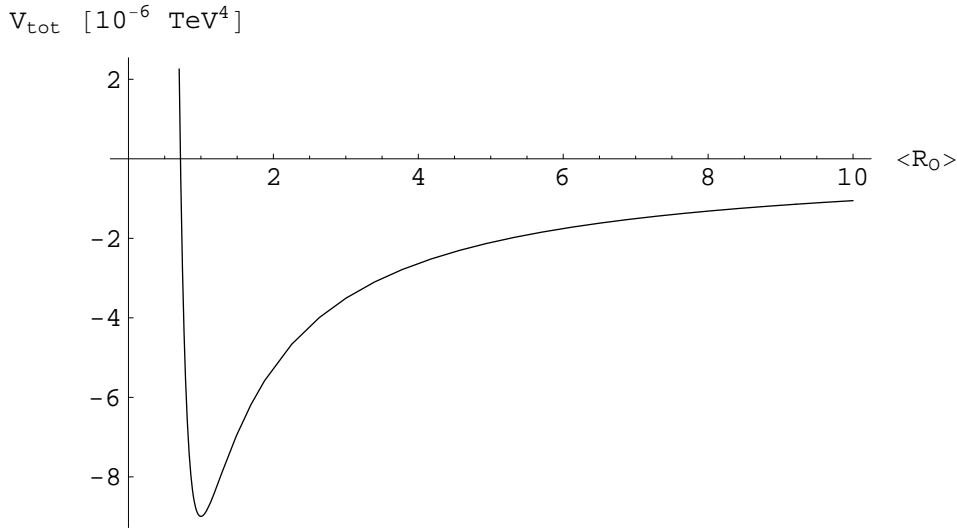, width=.75 \linewidth}
        \caption{\small{The total potential for the parameters given in (\ref{valueforparnh}) and
         for $m_4=0.175 {\rm ~TeV}$.
         The minimum appears  at the point $ \langle R_0\rangle =1$.} \label{w2}}
        \end{center}
\end{figure}

The fermionic contribution to the one-loop effective potential reads:
           \begin{equation}
           V^{\rm 1-loop}_{f}=-4\left(V^{(\infty)}_{f}+V^{(R)}_{f}\right)\,,
           \end{equation}
           where
            \begin{eqnarray} \label{hsmf}
V^{(\infty)}_{f}&=& e^{\frac{3}{2}\alpha \langle r\rangle}\frac{\rho }{60\pi}|m_{f}|^{5}\,,\nonumber\\
V^{(R)}_{f}&=&-e^{-6\alpha \langle
r\rangle}\frac{1}{64\pi^{6}\rho^{4}}\left[x_{f}^{2}{\rm
Li_{3}}\left(e^{-x_{f}}\right)+ 3 x_{f}{\rm
Li_{4}}\left(e^{-x_{f}}\right)+3{\rm
Li_{5}}\left(e^{-x_{f}}\right)\right]\,.
           \end{eqnarray}
We have defined $m_{f}=e^{-\frac{1}{2}\alpha \langle r\rangle}m_{4}$ and
$x_{f}=e^{\frac{3}{2}\alpha \langle r\rangle}2\pi\rho|m_{f}|$.
The total one-loop potential including all contributions reads
            \begin{equation}
            V^{\rm 1-loop}_{\rm tot}=V^{\rm 1-loop}_{v}+V^{\rm 1-loop}_{f}\,.
            \end{equation}
We have again analysed the effective potential as a function of
$\langle r\rangle$. Numerical calculations have been done for the
following choice of parameters: \beq \kappa=0.1{\rm ~TeV}\,,\quad
M_4=2\times 10^{15}{\rm ~TeV} \label{valueforparnh} \eeq and for
various values of the fermion mass. We have chosen the parameter
$\rho$ in such a way that the minimum of the complete  potential
appears at the point $\langle r\rangle\ll M_{\rm Pl}$, which
implies $R_0\approx 1$ (see Fig.~\ref{w2}). In such a case the
physical radius of the fifth dimension is given by the parameter
$\rho$. We have found the mass of the radion in the form
            \begin{equation}
            m^{2}_{r}=\frac{\partial^{2}V^{\rm 1-loop}_{\rm tot}}{\partial \langle r\rangle^{2}
            }\bigg|_{\rm min},
            \end{equation}
and the value of the scalar potential at the minimum $\Lambda_4$, which is
the cosmological constant. We have displayed the results in Table~\ref{tab2}.
One can easily find the following approximate relations between the mass of the
fermion and the other physical parameters
        \begin{equation}
        \rho^{-1}=c_{\rho}m_4, \quad m_r=c_m \frac{m_4^2}{M_{\rm Pl}}, \quad \Lambda_4=-c_{\Lambda} m_4^4\, ,
        \end{equation}
        where
\begin{equation}
        c_{\rho}=1.9, \quad c_m=5.2\times 10^{-2}, \quad c_{\Lambda}=9.6\times 10^{-3}\,.\label{relat}
\end{equation}
\begin{table}[h!]
        \begin{center}
            \begin{tabular}{|c|c|c|c|}\hline
              $m_4$ [TeV]&$\rho^{-1}$ [TeV]&$m_{r}$ [eV]&$\Lambda_4$ [TeV$^4$]\\\hline
              0.08\phantom{0}&0.15&$6.7\times 10^{-7}$&$-3.93\times 10^{-7}$\\\hline
              0.175&0.33&          $3.2\times 10^{-6}$&$-8.99\times 10^{-6}$\\\hline
              0.35\phantom{0}&0.66&$1.3\times 10^{-5}$&$-1.44\times 10^{-4}$\\\hline
              0.7\phantom{00}&1.32&$5.1\times 10^{-5}$&$-2.30\times 10^{-3}$\\\hline
              1.4\phantom{00}&2.65&$2.1\times 10^{-4}$&$-3.69\times 10^{-2}$\\\hline
              2.8\phantom{00}&5.30&$8.2\times 10^{-4}$&$-5.90\times 10^{-1}$\\\hline
            \end{tabular}
\caption{Radion masses together with the scale of the extra dimension $\rho^{-1}$  and the
resulting cosmological constant,  obtained for different input fermion masses $m_4$. }
\label{tab2}
        \end{center}
\end{table}
It is seen from the table and relations (\ref{relat}) that the dependence of
the radion mass on the input bulk fermion mass,
$m_5=\sqrt{2\pi\rho}m_4$, is quite strong. The result is the variation of the
radion mass between $6.7\times 10^{-7}$ and $8.2\times 10^{-4}$~eV.
Notice that the value of the cosmological constant that we have obtained is much larger than
cosmological constraints. However, one can cancel it by the renormalization counterterms.
To obtain a constant counterterm in the 4d
theory, after the Weyl rescaling, the following corrections can be added to the 5d action:
    \begin{equation}
    \delta S=\int d^5x \sqrt{G} \sqrt{-G_{55}} \delta \Lambda + \int d^4x \sqrt{-g}G_{55}\delta \lambda_0\Big|_{y=0}
    + \int d^4x \sqrt{-g}G_{55}\delta \lambda_\pi\Big|_{y=\pi\rho}\,,
    \end{equation}
where the first term spoils 5d covariance in the bulk, but is acceptable
from the 4d point of view (also, it is considered here as a one-loop-order counterterm).
These counterterms can be used to make the 4d one-loop cosmological constant
vanish without violating the 4d Lorentz invariance and, more importantly,
without destabilizing the scalar potential for the radion.

It can be seen that the presence of the 5d bulk mass term for the
fermions is crucial for the stabilization. The minimum at a finite
value of the radion disappears when $m_5$ approaches zero (this is
the decompactification limit, and the radion expectation value
runs away toward infinity). The point is that the dependence of
the tree-level fermionic mass term on the radion is different from
that of the KK mass terms, and the presence of the minimum is the
result of the interplay between the terms denoted as
$V_{f}^{(\infty)}$ and $V_{f}^{(R)}$ in (\ref{hsmf}), the first of
which depends on the tree-level fermionic mass, the second on the
KK masses.

\section{Summary}
\label{sum}

We have discussed the stabilization of the scalar sector including the radion,
in the QED-like gauge model with one universal extra dimension; with gauge symmetry
broken by the 5d Higgs mechanism and in the case where the breaking occurs because of the boundary conditions
imposed on the gauge fields.
The stabilization is due to the fermionic contribution to the effective potential.
In fact, for the stabilization  to take place, the bosonic contribution must be balanced by the fermionic one,
hence the scales of these two cannot differ too much. However, one does not need (softly broken)
supersymmetry to achieve the stabilization: it can be arranged in models born in
universal extra dimensions for a reasonably wide range of couplings and mass scales.
One does not need complicated models or unreasonable fine-tunings; even the simple
QED-like set-up is sufficient.
We expect the generic features of our mechanism to hold also in the case of (broken) supersymmetry,
even in the presence of a larger number of  moduli
fields (see also \cite{vonGersdorff:2003rq}).

It can be seen that the presence of the 5d bulk mass term for the
fermions is crucial for the stabilization. For instance, in the
Higgs model discussed in Section~\ref{radcor}, the minimum at
finite values of the fields in the radion--scalar hyperplane
disappears when $g_5$ (so consequently the mass of the zero-mode
fermion vanishes) approaches zero (this is the decompactification
limit, and the radion vacuum expectation value runs away toward
infinity). The point is that the dependence of the tree-level
fermionic mass term on the radion is different from that of the KK
mass terms, and the presence of the minimum is the result of the
interplay between the terms denoted as $V_{f}^{(\infty)}$ and
$V_{f}^{(R)}$ in (\ref{hmf}), the first of which depends on the
tree-level fermionic mass, the second on the KK masses. The
situation is very similar in the Higgsless case, for which the
relevant formula is (\ref{hsmf}).

One may also consider localized brane mass terms for the fermions
of the form $G_{55} \delta(x^5-x^{5}_{b}) m_b \overline{\psi}
\lambda$. However, these terms play the role of sources in the
equations of motion, and they are cancelled by discontinuities in
the bulk fermionic configurations. Their role is to impose
boundary conditions on the fields, hence they affect the
quantization of the masses of the KK modes. This effect on its own
does not create a minimum: the bulk terms described above are
still needed.

It is interesting to watch correlation between the various
physical parameters that arise upon the stabilization of the
scalar sector. For a Higgs mass larger than $0.26$~TeV, we observe
that there appears an instability in the effective potential in
the direction of the radion - its vacuum expectation value runs
away to infinity (decompactification limit). It is interesting to
note that $m_H \simeq 0.204$~TeV is  the electroweak 95\% CL upper
bound on the Higgs boson mass. Therefore even though our toy model
does not reflect all the features of the real 5d SM, it
nevertheless favours the range of Higgs boson masses that is also
anticipated by the one-loop predictions of the Standard Model. It
turns out that, for parameter values adopted here for the Higgs
case, the radion run-away is the primary instability in the model,
not the large-$h$ instability discussed in \cite{Bucci:2003fk}.

It is also interesting to note that the cosmological constant may be cancelled by suitable counterterms,
in such a way that stabilization of scalars is not affected.
\vspace{1cm}

\centerline{\Large \bf Acknowledgements}

\vspace*{.5cm}

B.G. is supported in part by the State Committee for Scientific
Research (Poland) under grant 1~P03B~078~26 in the period 2004--2006.
Z.L. thanks the Theory Division at CERN for hospitality. \noindent This work  was
partially supported  by the EC Contract HPRN-CT-2000-00152 for
the years 2000--2004, by the Polish State Committee for Scientific
Research grants KBN 2P03B 001 25 (Z.L.) and by KBN 2P03B 124 25
(R.M.), and by POLONIUM 2004.

\vspace*{.5cm}

\newpage
\noindent{\Large \bf Appendix }
\setcounter{equation}{0}
\renewcommand{\theequation}{A.\arabic{equation}}
\vspace{0.3cm}

Here we provide some details of the dimensional reduction (in the Higgs case)
and calculation of the effective potential
generated by a tower of KK modes.

The KK expansion of the fields living on $S^{1}/Z_2$ gives
\begin{eqnarray}
A^{\mu}(x,y)&=&\frac{1}{\sqrt{2\pi\rho}}\left[A_{0}^{\mu}(x)+\sqrt{2}
\sum_{n=1}^{\infty}A_{n}^{\mu}(x)\cos(m_{n}y)\right] \,,\nonumber\\
A^{5}(x,y)&=&\frac{1}{\sqrt{\pi\rho}}\sum_{n=1}^{\infty}A_{n}^{5}(x)\sin(m_{n}y) \,,\nonumber\\
\phi(x,y)&=&\frac{1}{\sqrt{2\pi\rho}}\left[
\phi_{0}(x)+\sqrt{2}\sum_{n=1}^{\infty}\phi_{n}(x)\cos(m_{n}y)\right] \,,\nonumber\\
 \psi(x,y)&=&\frac{1}{\sqrt{2\pi\rho}}\left[\psi_{R0}(x)+\sqrt{2}\sum_{n=1}^{\infty}\left[\psi_{Rn}(x)\cos(m_{n}y)+
\psi_{Ln}(x)\sin(m_{n}y)\right]\right]
\,,\nonumber\\\lambda(x,y)&=&\frac{1}{\sqrt{2\pi\rho}}\left[\lambda_{L0}(x)+
\sqrt{2}\sum_{n=1}^{\infty}\left[\lambda_{Ln}(x)\cos(m_{n}y)+\lambda_{Rn}(x)\sin(m_{n}y)\right]\right]
\,, \label{kk}
\end{eqnarray}
where $m_{n}=2\pi n/L$.

Expanding the 4d Lagrangian around
$h_0 \rightarrow h_0 + \langle h\rangle $, $r\rightarrow r+\langle r\rangle$,
the following scalar mass terms are obtained in the Landau gauge:
     \begin{eqnarray}
        m^{2}_{h_{0}}&=& e^{-\alpha \langle r\rangle}\left(-\mu^{2}+3\lambda_{4}\langle h\rangle^{2}\right)\,,\nonumber\\
        m^{2}_{\chi_{0}}&=&e^{-\alpha \langle r\rangle}\left(-\mu^{2}+\lambda_{4}\langle h\rangle^{2}\right) \,,\nonumber\\
        m^{2}_{h_{n}}&=&e^{-\alpha \langle r\rangle}\left(-\mu^{2}+3\lambda_{4}\langle h\rangle^{2}+e^{-2\alpha \langle r\rangle}m_{n}^{2}\right)\,,\nonumber\\
        m^{2}_{\chi_{n}}&=&e^{-\alpha \langle r\rangle}\left(-\mu^{2}+\lambda_{4}\langle h\rangle^{2}+e^{-2\alpha \langle r\rangle}m_{n}^{2}\right) \,,\nonumber\\
        m^{2}_{r}&=&\alpha^{2}e^{-\alpha \langle r\rangle}\left(-\frac{1}{2}\mu^{2}\langle h\rangle^{2}+\frac{1}{4}\lambda_{4}\langle h\rangle^{4}+\frac{\mu^{4}}{4\lambda_4}\right) \,,\nonumber\\
        m^{2}_{r\; h_{0}}&=&-\alpha e^{-\alpha \langle r\rangle}\left(-\mu^{2}\langle h\rangle+\lambda_{4}\langle h\rangle^{3}\right) \,,\nonumber\\
        m^{2}_{A_{5n}}&=&e^{-\alpha \langle r\rangle}e_{4}^{2}\langle h\rangle^{2} \,,\nonumber\\
         m^{2}_{A_{5n}\;\chi_{n}}&=&-e^{-2\alpha \langle
r\rangle}e_{4}\langle h\rangle m_{n} \,, \label{mass}
\end{eqnarray}
where $\alpha$, $\lambda_{4}$, $e_{4}$ are defined in the main text.

For vector bosons the following mass terms are obtained
\begin{eqnarray}~\label{massvb}
m^{2}_{A_{\mu n}}&=& e^{-\alpha \langle
r\rangle}\left(e_{4}^{2}\langle h\rangle^{2}+e^{-2\alpha \langle
r\rangle}m_{n}^{2}\right)
\,,\nonumber\\
m^{2}_{A_{\mu 0}}&=&e^{-\alpha \langle r\rangle}e_{4}^{2}\langle
h\rangle^{2} \,.
\end{eqnarray}

The masses of the fermions are given by
\begin{eqnarray} \label{massf}
m_{\psi_{n}}&=& -e^{-\frac{3}{2}\alpha \langle r\rangle}m_{n}
\,,\nonumber\\m_{\lambda_{n}} &=& e^{-\frac{3}{2}\alpha \langle
r\rangle}m_{n} \,,\nonumber\\m_{\psi_{n}\;\lambda_{n}}
&=& -e^{-\frac{1}{2}\alpha \langle r\rangle}\frac{g_{4}}{\sqrt{2}}\langle h\rangle \,,\nonumber\\
m_{\psi_{0R}\;\lambda_{0L}} &=& -e^{-\frac{1}{2}\alpha \langle
r\rangle}\frac{g_{4}}{\sqrt{2}}\langle h\rangle \,,
\end{eqnarray}
where $g_{4}=g_{5}/\sqrt{2\pi\rho}$.

For the purpose of this paper we have adopted the
regularization
scheme worked out by Delgado et al. (DPQ, see \cite{Delgado:1998qr}) to
     compute the contribution of the KK tower to the effective potential.
Let us briefly recall the basic result obtain by DPQ.\\
Starting from the generic formula
\begin{equation}
V(\phi)=\frac{1}{2}\int\frac{d^{4}p}{(2 \pi)^{4}}\sum_{0}^{\infty}\log{[l^2 E^2+n^2 \pi^2]}\,,
\label{dpq}
\end{equation}
where $E^2\equiv p^2+m^2(\phi)$, $m^2(\phi)$ are the background-field-dependent
mass squared of the KK modes and $l=\pi \rho$. With the help of
the $\overline{MS}$ renormalization scheme one obtains
\begin{equation}
V=\frac{1}{2}(V^{(\infty)}+V^{(R)}+V^{0})
\end{equation}
where
\begin{eqnarray}
V^{(\infty)} & = & \frac{ \rho}{60 \pi}m^5(\phi)\,,\nonumber \\
V^{0} & = & \frac{1}{64\pi^{2}}m^{4}
(\phi)\left[\log\left(\frac{m^{2}(\phi)}{k^{2}}\right)-\frac{3}{2}\right]\,,\nonumber\\
V^{(R)} & = & -\frac{1}{64 \pi^{6} \rho^{4}}(x^2 {\rm
Li}_{3}(e^{-x})+3x {\rm Li}_{4}(e^{-x})+3 {\rm Li}_{5}(e^{-x}))\,.
\end{eqnarray}
In the above the $x$ is given by $x=2 \pi \rho \sqrt{m^2(\phi)}$,
$\kappa$ is the renormalization scale, and ${\rm
Li}_{n}(x)=\sum_{s=1}^{\infty}\frac{x^{s}}{s^{n}}$ is the
polylogarithm function.

\vspace*{0.3cm}


\end{document}